\documentclass{article}

\usepackage{PRIMEarxiv}

\usepackage[utf8]{inputenc} 
\usepackage[T1]{fontenc}    
\usepackage{hyperref}       
\usepackage{url}            
\usepackage{booktabs}       
\usepackage{amsfonts}       
\usepackage{nicefrac}       
\usepackage{microtype}      
\usepackage{lipsum}
\usepackage{fancyhdr}       
\usepackage{graphicx}       
\graphicspath{{media/}}     
\usepackage{multirow}   
\usepackage{makecell}   
\usepackage{caption}    
\usepackage{amsmath}    
\usepackage{tabularx}

\title{Morpheus: Lightweight RTT Prediction for Performance\allowbreak -Aware Load Balancing
\thanks{This manuscript is a preprint and has been submitted for possible publication. It has not undergone peer review.}
}

\author{
  Panagiotis Giannakopoulos \\
  Eindhoven University of Technology \\
  Eindhoven, The Netherlands\\
  \texttt{p.giannakopoulos@tue.nl} \\
   \And
   Bart van Knippenberg \\
  Thermo Fisher Scientific \\
  Eindhoven, The Netherlands\\
  \texttt{bart.vanknippenberg@thermofisher.com} \\
  \And
   Kishor Chandra Joshi \\
  Eindhoven University of Technology \\
  Eindhoven, The Netherlands\\
  \texttt{k.c.joshi@tue.nl} \\
  \And
   Nicola Calabretta \\
  Eindhoven University of Technology \\
  Eindhoven, The Netherlands\\
  \texttt{n.calabretta@tue.nl} \\
  \And
   George Exarchakos \\
  Eindhoven University of Technology \\
  Eindhoven, The Netherlands\\
  \texttt{g.exarchakos@tue.nl} \\
}

\begin{document}
\maketitle

\begin{abstract}
Distributed applications increasingly demand low end-to-end latency, especially in edge and cloud environments where co-located workloads contend for limited resources. Traditional load-balancing strategies are typically reactive and rely on outdated or coarse-grained metrics, often leading to suboptimal routing decisions and increased tail latencies. This paper investigates the use of round-trip time (RTT) predictors to enhance request routing by anticipating application latency. We develop lightweight and accurate RTT predictors that are trained on time-series monitoring data collected from a Kubernetes-managed GPU cluster. By leveraging a reduced set of highly correlated monitoring metrics, our approach maintains low overhead while remaining adaptable to diverse co-location scenarios and heterogeneous hardware. The predictors achieve up to 95\% accuracy while keeping the prediction delay within 10\% of the application RTT. In addition, we identify the minimum prediction accuracy threshold and key system-level factors required to ensure effective predictor deployment in resource-constrained clusters. Simulation-based evaluation demonstrates that performance-aware load balancing can significantly reduce application RTT and minimize resource waste. These results highlight the feasibility of integrating predictive load balancing into future production systems.
\end{abstract}


\section{Introduction} \label{sec:introduction}
Modern distributed applications and services, such as web microservices, real-time multimedia streaming, and Open Radio Access Network (O-RAN) applications, demand low end-to-end latency~\cite{polese_understanding_2023}. These services, including our focus on electron microscopy (EM) workflows, operate in environments where load balancers play a crucial role by routing requests to application instances and making rapid decisions that directly impact the user experience. Traditional load-balancing strategies typically rely on aggregate metrics (e.g., server load) measured at fixed intervals. Although these reactive methods are straightforward to implement and monitor overall performance trends, they rely on coarse-grained or outdated data, often resulting in suboptimal routing decisions, degraded performance, and increased tail latencies~\cite{wydrowski_load_2024}.

To reduce operational costs, applications are co-located in edge and cloud computing environments. However, co-location increases competition for limited resources, introducing interference and execution-time variability~\cite{paasivaara_predictable_2020}. Furthermore, diverse hardware configurations and software stacks across edge/cloud nodes deteriorate performance variability, leading to significant differences in application performance between runs and servers~\cite{maricq_taming_nodate,duplyakin_studying_2019}. Thus, meticulous management of computing and network resources is essential to achieve predictable performance~\cite{fu_progress-based_2019}, maximize resource utilization~\cite{townend_invited_2019}, and automate infrastructure management.

Recent advancements in machine learning and time-series modeling have made it possible to accurately predict round-trip time (RTT). These predictors often use system monitoring metrics, such as CPU utilization~\cite{mohamed_end--end_2021}, or system characteristics, like CPU configuration~\cite{wang_predicting_2018}. However, most of the existing approaches have significant limitations. They depend on a small, fixed set of coarse-grained metrics and do not adapt well to changing co-location scenarios or heterogeneous hardware. Despite their promise, prediction-based load balancing is rarely adopted in production systems. One key obstacle is that accurate RTT predictions depend on complex machine learning workflows that require ongoing retraining and fine-tuning, leading to additional computational and communication overhead. Integrating these predictive models into high-speed load balancers, while maintaining throughput, responsiveness, and reliability, remains a major challenge.

To address this challenge, we examine performance predictability within an EM workflow, specifically focusing on the Single Particle Analysis (SPA) acquisition and processing pipeline. SPA workflows operate at near-real-time instrument control with sub-second requirements, demanding heterogeneous resources, including CPU, GPU, and network. Typical EM setups experience fluctuating resource usage, with bursts of high activity followed by idle periods. Deploying SPA applications on shared, distributed edge infrastructure allows resource pooling from multiple EM systems, dynamically reallocating resources to smooth usage spikes and reduce idle times. In this context, proactive load balancing guided by predictive RTT models can efficiently distribute workloads across edge nodes, improving resource utilization and ensuring reliable, low-latency execution. This approach enables a scalable and cost-effective SPA infrastructure while meeting strict requirements for data privacy, latency, and reliability.

In this work, we present Morpheus, a framework that automatically builds lightweight RTT predictors. Experiments are conducted on a Kubernetes-managed GPU cluster, with monitoring metrics collected at 200ms intervals using Prometheus. The predictors run alongside applications, training and evaluating multiple models to select the optimal one based on accuracy and prediction delay. By leveraging a reduced set of highly correlated monitoring metrics, they minimize computational overhead while maintaining accuracy across diverse co-location scenarios.

We evaluate the predictors in a real runtime system, measuring accuracy, overhead, and prediction speed across heterogeneous servers. The results show high accuracy (up to 95\%) with low prediction overhead (less than 10\% of the RTT) and identify bottlenecks that limit prediction speed, allowing targeted optimizations for time-sensitive applications. Building on these runtime results, we examine their impact on load balancing through simulation, demonstrating substantial improvements in request completion times and identifying critical accuracy thresholds beyond which further gains become negligible. The main contributions of this work are:
\begin{itemize}
    \item  A practical methodology for designing and implementing RTT predictors that operate alongside applications and progressively learn their performance characteristics
    \item A simulation study demonstrating performance\allowbreak -aware load balancing across varying accuracy, application instance counts, and cluster heterogeneity.
    \item Optimization insights to improve predictor accuracy and efficiency in edge clusters.
\end{itemize}

The paper is organized as follows: Section~\ref{sec:related_work} reviews related work; Section~\ref{sec:predictor_architecture} describes the predictor architecture and Section~\ref{sec:set_up} the experimental setup. Sections~\ref{sec:predictor_results} and~\ref{sec:simulation_results} present the predictor evaluation and load-balancing results. Section~\ref{sec:lessons} summarizes lessons learned, and Section~\ref{sec:conclusions} concludes with future directions.

\section{Related work} \label{sec:related_work}

Applications in resource clusters often experience unpredictable performance due to shared resources, hardware aging, and frequent system updates~\cite{Benchmarking,ballani2011towards}. Previous studies have analyzed the sources of this variability, including differences in hardware, software, and deployment~\cite{chunduri_run--run_2017,maricq_taming_nodate,duplyakin_studying_2019,traini_how_2021,paasivaara_predictable_2020,laaber_evaluating_2024}. Although frameworks such as CONFIRM~\cite{maricq_taming_nodate} and various observability tools~\cite{reichelt_automated_2022,sukhija_event_2020,mart_observability_2020} can diagnose or quantify performance variation and detect anomalies, these approaches are typically reactive and lack predictive capabilities. Trace-based microservices methods~\cite{qiu_causality_2020,meng_detecting_2021,wu_microdiag_2021} provide valuable post hoc analysis, but do not enable a proactive response. Building on these foundations, our work incorporates identified variability sources into the design of efficient, real-time predictors.

Learning-based performance prediction has gained traction in recent years. Zhao et al.~\cite{zhao_cloud_2021} applied gradient-boosted models to detect performance changes due to system updates, but focused on long-term trends and ignored co-location effects. Mohamed et al.~\cite{mohamed_end--end_2021} used machine learning with a limited set of metrics and SHAP~\cite{lundberg_unified_2017} for interpretability, although its computational overhead limits real-time use. We instead adopt perfCorrelate~\cite{giannakopoulos_perfcorrelate_2025}, a lightweight correlation-based method for metric selection that enables fast predictions. Some works estimate the remaining completion time using application-level progress signals~\cite{fu_progress-based_2019}, which require code instrumentation and are not suitable for black-box systems. Our approach predicts performance ahead of execution using monitoring data that are already collected from the deployed monitoring system. Existing cluster-level scheduling solutions~\cite{delimitrou_quasar_2014,dauwe_hpc_2016,townend_invited_2019} improve efficiency in deployment time but do not address per-request decision-making. In contrast, our method enables fine-grained real-time traffic steering among replicas.

Traditional load balancing policies, such as round-robin and least-connections, are easy to implement, but static and unresponsive to dynamic changes in the system~\cite{alankar_experimental_2020,laukka_load_2023}. Recent context-aware and adaptive strategies, including cloud-based LSTM forecasting~\cite{chen_context-aware_2023}, adaptive weighting~\cite{michaelis_l3_2024}, and performance\allowbreak -driven frameworks~\cite{aslanpour_load_2024}, offer more flexibility but rely on static profiles or react after overloads rather than preventing them. Similarly, Prequal~\cite{wydrowski_load_2024} estimates server quality reactively without anticipating future performance. Collaborative and distributed approaches, such as BLOC~\cite{bhattacharya_bloc_2022} and federated ML scheduling~\cite{kathole_novel_2025}, use feedback or global models but introduce considerable latency and overhead, limiting real-time adaptation at the request level. 

In summary, although previous work has advanced the understanding of performance variability, anomaly detection, and scheduling, fast per-request performance prediction for performance\allowbreak -aware load balancing remains an open challenge. Our work addresses this gap by designing lightweight models that use live monitoring data to estimate RTT in real time, enabling dynamic, performance\allowbreak -aware request routing for time-sensitive applications.

\section{Predictor architecture} \label{sec:predictor_architecture}
\begin{figure}
\centerline{\includegraphics[width=0.5\columnwidth]{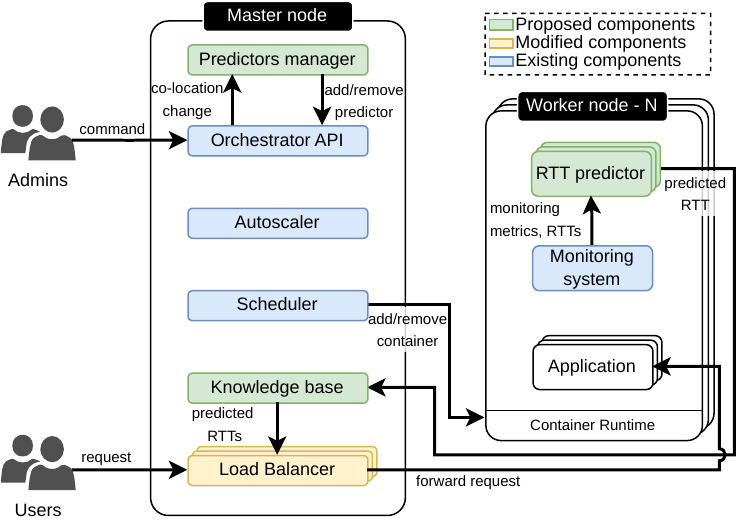}}
\caption{High-level architecture for integrating RTT predictors into the load balancer of a resource cluster.}  
\label{fig:high_level_architecture}
\end{figure}
The proposed run-time predictors operate alongside the target application and continuously learn its performance behavior over time. Figure~\ref{fig:high_level_architecture} provides a high-level overview of how RTT predictors are integrated into the load balancer. Application performance can vary significantly between different server configurations, such as due to differences in CPU models~\cite{chunduri_run--run_2017,maricq_taming_nodate,duplyakin_studying_2019}. Therefore, a separate predictor is deployed for each unique combination of application and server within the cluster. Each predictor collects RTT measurements along with the corresponding monitoring metrics for every task execution using the monitoring system. These data are used to train and update machine learning models that estimate the RTT for future tasks. In this context, RTT is defined as the time interval between when a request enters the cluster gateway ($t_{start}$) and when the corresponding response exits the gateway ($t_{end}$). This definition captures in-cluster performance by excluding network transmission delays between the client and the gateway. Each task corresponds to a single request-response cycle initiated by the user. 

\begin{figure*}[h]
\centerline{\includegraphics[width=1\columnwidth]{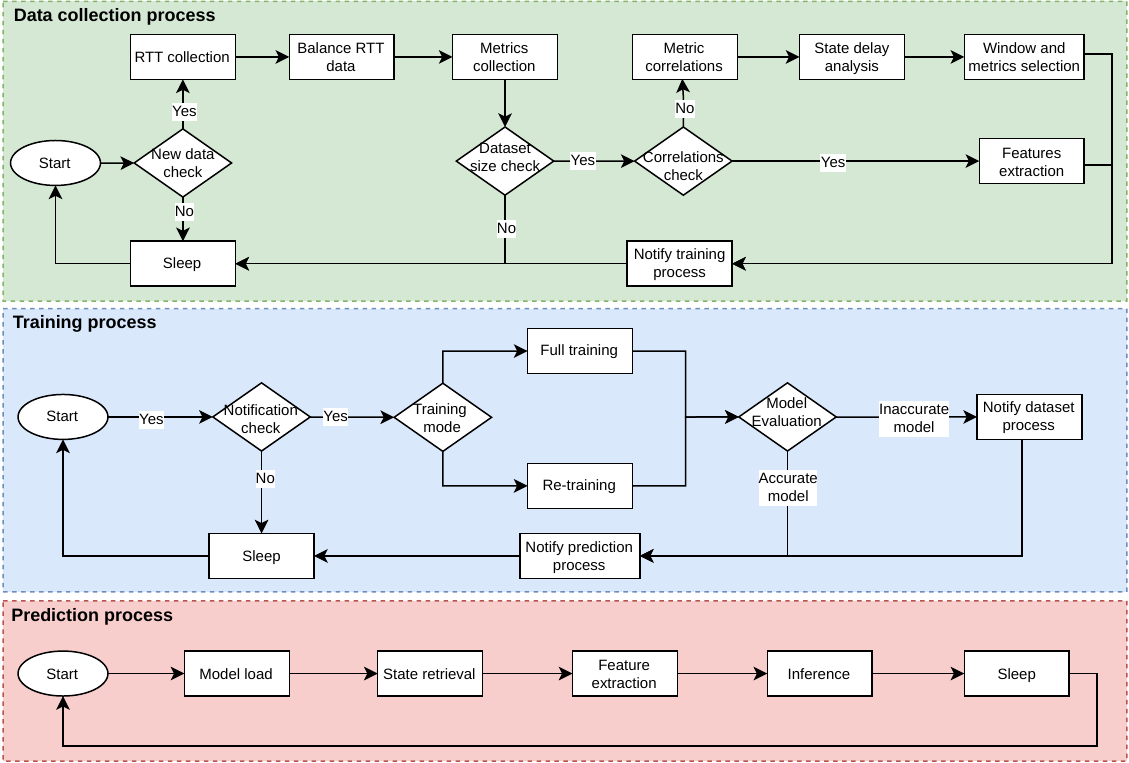}}
\caption{Overview of the RTT prediction methodology.} 
\label{fig:methodology}
\end{figure*}

Predictions are stored in a database called the knowledge base and later used by each application load balancer to guide placement decisions. The Prediction Manager monitors application activity within the cluster, deploying an RTT predictor when an application runs on a server for the first time, re-enabling it if previously deployed, and pausing it when all instances are removed. To ensure accurate RTT prediction for new tasks, the predictors of Morpheus run three parallel processes: data collection, training, and prediction, as shown in Figure~\ref{fig:methodology}.

\textbf{Data Collection:}  
This process collects RTT and monitoring metrics for each completed task using the monitoring system. For each task, it collects monitoring data from the period just before task submission (referred to as observation window), along with the corresponding RTT value. Correlation analysis, through perfCorrelate~\cite{giannakopoulos_perfcorrelate_2025}, is performed to evaluate the relationship between each monitoring metric and the RTT. This is done using several correlation algorithms and different observation windows. As a result, the process identifies both the most relevant metrics and the best observation window to capture system state. The size of the retrieved system state is determined by the number of selected metrics and the length of the observation window. To ensure that the system state does not introduce unacceptable delays during predictions, a strict time limit is imposed on its retrieval. The dataset is stored in two forms: raw time-series data and feature-based representations, such as mean or standard deviation, to support different types of models.

\textbf{Training:}  
As the dataset is updated, the process creates or updates machine learning models using the latest monitoring metrics and RTTs. These metrics are prepared in two forms to accommodate different types of models. For models specialized in time-series data, such as recurrent neural networks, the raw time-series of monitoring metrics are used as input. For models that do not require sequential data, such as XGBoost, statistical features (for example, mean and standard deviation) are extracted from each monitoring metric. The process evaluates various model types with both data representations and selects the one that provides the highest prediction accuracy while meeting the required inference time constraints.  

\textbf{Prediction:}  
For each new task, the process loads the best model trained by the model training process. The process retrieves the required recent data from the system state, extracts features if necessary, and uses the model to estimate the RTT for the upcoming task. The predicted RTT is stored in the knowledge base for further use by the load balancer.

\subsection{Data collection process}
The data collection process runs at regular intervals, repeating every five minutes. Each cycle includes a sequence of steps that correspond to the components shown in the green panel of Figure~\ref{fig:methodology}. The steps are the following:


\underline{New data check:} The process checks whether the monitoring system has recorded new task records since the last cycle. If no new data are found, the process enters a sleep state until the next scheduled run.

\underline{RTT collection}: In case new tasks exist, they are retrieved from the monitoring system. Specifically, for each task, the ($t_{start}$), ($t_{end}$) and RTT are stored. 

\underline{Balance RTT data:}
In practice, tasks can arrive at a high rate, causing the dataset to grow rapidly as new data are collected. When the data distribution is skewed, many of these new samples can be duplicates or highly similar to existing ones, leading to unnecessary redundancy in the dataset. Storing all task records and loading them into memory becomes impractical, particularly for servers with limited resources. This challenge is even more pronounced when dealing with resource usage monitoring metrics. For example, if we store 60 seconds of historical data prior to each task submission for 500 metrics, with each metric sampled at 5 values per second (i.e., every 200\,ms), this results in $500 \times 5 \times 60 = 150{,}000$ data points per task. 

Additionally, if the RTT distribution in the dataset is non-uniform, the model may become biased toward the most common RTT values, which reduces prediction accuracy~\cite{he_learning_2009}. To address both memory limitations and the risk of model bias, we construct a training dataset that is compact and preserves the RTT distribution as uniformly as possible. This is achieved using a dynamic binning strategy, which selectively retains only those samples that help preserve a balanced RTT distribution. The details of how this strategy operates depend on whether an initial dataset is already present:

\textit{Case 1 (no existing training data):}
In the initial phase, when no prior dataset exists, all collected RTT samples are retained without any filtering or subsampling. This approach ensures that potentially valuable information is not discarded at the outset, preserving the complete diversity and variability of the observed RTT values. Keeping the full set of initial data is especially practical in this context because data acquisition occurs at a moderate frequency (e.g., every 5 minutes), which means that the dataset size remains small and manageable during early collection periods. By retaining all initial samples, the method provides a comprehensive and unbiased foundation for defining future dataset structure and balancing criteria as new data become available.

\textit{Case 2 (existing training data):}  
When new RTT samples arrive, the bin boundaries are recalculated using the combined RTT values from both the existing dataset and the new samples. Specifically, the set \( S \) consists of all RTT values from both sources, and the bin width is computed using the Freedman–Diaconis rule~\cite{freedman_histogram_1981}:
\begin{equation}
    h = \frac{2 \cdot \text{IQR}(S)}{N^{1/3}},
\end{equation}
where \( \text{IQR}(S) = Q_{75}(S) - Q_{25}(S) \) and \( N \) is the total number of RTT samples in \( S \). The number of bins \( l \) and their upper boundaries \( b_i \) are then:
\begin{equation}
l = \left\lceil \frac{\max(S) - \min(S)}{h} \right\rceil,
\quad
b_i = \min(S) + i \cdot h, \quad i = 1 \cdots l
\end{equation}
Each new RTT sample is then assigned to a bin, and retained only if the number of existing samples in that bin is less than the maximum bin count, \( C_{\text{max}} \), across all bins. If more new samples arrive for a bin than its available capacity, a random subset of new samples is selected to fill the gap. The remaining capacity for bin \( i \) is given by:
\begin{equation}
    \text{gap}_i = C_{\text{max}} - c_i,
\end{equation}
where \( c_i \) is the existing sample count in bin \( i \). New samples are added to their bin up to this available capacity. If no bin has remaining capacity for any new sample, a single new sample is retained at random to ensure the dataset can continue to evolve.

In principle, RTT distribution uniformity could be further improved by removing samples from overrepresented bins in the existing dataset. However, this would also require removing the associated monitoring metrics for each RTT sample. Experimental measurements show that each RTT entry is about 77B, while the corresponding metrics total over 500kB due to 294 metric lines of roughly 1.8kB each. Because the monitoring dataset is orders of magnitude larger, removing samples would require extensive coordination and significantly increase both memory usage and processing time. Therefore, the proposed method focuses on selectively adding new samples, providing a practical balance between improving uniformity and minimizing overhead.

\underline{Metrics collection:} For each new task in the balanced dataset, the process retrieves the monitoring metrics from a 60-second window that precedes each task submission. This window is chosen because it is long enough to capture short-term fluctuations and resource contention that can affect task performance, yet short enough to avoid including outdated or irrelevant system states. 

\underline{Dataset size check:} The process assesses whether the existing dataset has sufficient data points to capture performance variability. This is done through CONFIRM~\cite{maricq_taming_nodate}, which estimates the minimum number of repetitions needed to ensure that the median RTT is within $r\%$ of the true median with $\alpha\%$ confidence. This method is robust for non-normally distributed data~\cite{normality-check}. If the dataset contains a sufficient number of samples, the process continues with the subsequent steps. Otherwise, it skips the remaining steps for the current cycle and waits until the next one.

\underline{Correlations check:} This step checks whether the previously established correlations between monitoring metrics and RTT remain valid. Correlations are considered valid if they exist and no notification has been issued by the training process (more details in Section~\ref{sec:training_process}). If the correlations are valid, the process jumps to the \textit{feature extraction} step.

\underline{Metric correlations:} The correlation analysis is performed by invoking the \texttt{perfCorrelate} module~\cite{giannakopoulos_perfcorrelate_2025}. This module operates in three stages: (1) it extracts, for each monitoring metric in the form of a time-series, the most relevant statistical feature using \texttt{tsfresh}~\cite{christ_time_2018}, selecting the one with the highest correlation to RTT, (2) it eliminates redundant metrics to reduce dimensionality, and (3) it computes correlations between each remaining metric and RTT using multiple methods. The used correlations are: Pearson, Spearman, Kendall, Distance Correlation, and Maximum Information Coefficient (MIC). An overview of the correlation methods supported is provided in Table~\ref{tab:correlation_measures}. The absolute values of the correlation scores are used to assess the relevance of each metric. This approach ensures that, regardless of the correlation method, all scores are in the range between 0 and 1. The monitoring metrics values are correlated with RTT using observation windows of 1, 5, 20, and 60 seconds before the task is submitted.

For each monitoring metric, correlations are calculated using all available methods. The method that yields the highest absolute correlation score for a given metric is selected as its representative. This procedure is applied independently across all observation windows, resulting in a comprehensive set of correlation results indexed by metric, selected method, and observation window.

\begin{table}[h]
\centering
\footnotesize
\begin{tabularx}{\columnwidth}{l l X}
\toprule
\textbf{Method} & \textbf{Range} & \textbf{Description} \\
\midrule
Pearson   & $[-1,1]$ & Measures linear correlation between continuous variables. \\
Spearman  & $[-1,1]$ & Detects monotonic relationships using rank order of data. \\
Kendall   & $[-1,1]$ & Measures ordinal association based on concordant and discordant pairs. \\
Distance  & $[0,1]$  & Quantifies general dependence using pairwise distances between samples. \\
MIC       & $[0,1]$  & Identifies linear and nonlinear dependencies using mutual information. \\
\bottomrule
\end{tabularx}
\caption{Correlation methods used in \textit{perfCorrelate}.}
\label{tab:correlation_measures}
\end{table}

\underline{State delay analysis:} The overhead associated with retrieving time-series data from the monitoring system and extracting features using \texttt{tsfresh} is then measured. This includes the time required for state retrieval ($t_{\text{state}}^k$) and features calculation ($t_{\text{feature}}^k$) for each number $k$ of monitoring metrics. The process begins with the five most correlated metrics and incrementally adds the next five most correlated metrics in each step, continuing this procedure until all relevant features have been included.

\underline{Window and metrics selection:} The optimal combination of the observation window $w$, the correlation method $r$, and the number of metrics $k$ is selected to balance the predictive accuracy with the overhead of preparing the input, which includes both the retrieval of the state of the system and the extraction of features. Among all evaluated combinations of $w$, $r$ and $k$, only those that satisfy the predefined overhead constraint are considered for selection: 
\begin{equation}
    t_{\text{state}}^k + t_{\text{feature}}^k \leq \tau_{\text{prepare}} \cdot \mu_{\text{RTT}}
\end{equation}
where $\tau_{\text{prepare}}$ defines the maximum allowed fraction of the average RTT ($\mu_{\text{RTT}}$) that can be spent on input preparation. If no combination meets this condition, no predictor is trained for that application–node pair. In our work, we use $\tau_{\text{prepare}} = 9\%$. The optimal combination of $w$, $r$ and $k$, denoted by an asterisk, is the one that maximizes the total correlation with RTT:
\begin{equation}
(w^*, r^*, k^*) = \arg\max_{w, r, k} \left\{ \sum_{i=1}^{k} \left| \rho_{w,r}(m_i, \text{RTT}) \right| \right\},
\end{equation}
where $\rho_{w,r}(m_i, \text{RTT})$ is the correlation between monitoring metric $m_i$ and RTT, calculated over observation window $w$ using correlation method $r$.

\underline{Feature extraction:} In this step, monitoring metrics from newly collected samples are transformed into statistical features and added to the dataset. This transformation is only performed if metric correlations have not already been computed during the current cycle, as the correlation analysis step also handles feature extraction and dataset updates. Feature extraction is triggered only if required by the training process, specifically when the model expects input in the form of aggregated features rather than time-series data (i.e., non-sequential format). 

\underline{Training notification:}
Once the dataset is updated by either the \textit{Metrics Collection} or \textit{Feature Extraction} step, the training process is notified. After this notification, the data collection process pauses and waits until the next cycle begins.

\subsection{Model Training Process} \label{sec:training_process}
The model training process follows an event-driven manner, initiated whenever the model needs to be created or updated. This can occur either when new data arrives or when a decrease in the accuracy of the current model is detected. It consists of a sequence of steps corresponding to the components in the blue panel of Figure~\ref{fig:methodology}. These steps are as follows:

\begin{table*}[h]
\centering
\footnotesize
\begin{tabularx}{1\columnwidth}{l l l X}
\toprule
\textbf{Correlation Type} & \textbf{Dataset Size} & \textbf{Models} & \textbf{Justification} \\
\midrule

Pearson (Linear) 
& Any size 
& LR, XGB 
& Models linear relations efficiently. LR and XGB generalize well across dataset sizes with low complexity. \\
\midrule

Spearman / Kendall (Monotonic) 
& Any size 
& RF, XGB, SVM 
& Effective for monotonic trends. RF and XGB handle ordinal data; SVM captures non-linear boundaries. \\
\midrule


\multirow{3}{*}{Distance / MIC (Non-linear)} 
& Small ($<$1K) 
& XGB 
& XGB captures complex non-linear patterns in small datasets with minimal risk of overfitting. \\

& Medium (1K–10K) 
& XGB, FNN 
& Both XGB and shallow FNNs handle moderate-size data well, balancing complexity and accuracy. \\

& Large ($>$10K) 
& XGB, FNN, RNN, CNN 
& RNNs/CNNs learn intricate patterns; FNN and XGB still offer strong performance and efficiency. \\
\bottomrule
\end{tabularx}
\caption{Compact overview of suitable ML models based on correlation type and dataset size.}
\label{tab:combined-model-selection}
\end{table*}

\underline{Training mode:} The process chooses between two training modes: (1) \textit{full training}, where multiple model types are trained from scratch to find the best-performing model, and (2) \textit{re-training}, where the current model is updated with newly collected data. \textit{Full training} is used when there is no existing model or when the current model is no longer accurate enough. Otherwise, if the existing model still performs well, \textit{re-training} is performed.


\underline{Full training:}
In full training a broad set of candidate models is trained from scratch and evaluated. We categorize performance prediction models into two distinct groups: non-sequential and sequential models. This classification is based on the nature of the input data. Non-sequential models are suitable for structured datasets where the order of observations does not influence the outcome, while sequential models are designed to capture temporal dependencies and evolving patterns over time. This dual approach allows us to flexibly address both static and time-dependent behaviors in predicting application performance.

Each model in the respective categories receives input data \( X \), which represents the monitoring metrics either as extracted features or as time-series data, and produces output \( Y \), the predicted RTT for the target application. Both \( X \) and \( Y \) are normalized using MinMax scaling to a range between 0 and 1. Outliers are removed from the dataset (z-score $>$ 3) before further processing. The data is then divided into three subsets: 80\% for training, 10\% for validation, and 10\% for testing. The training set is used to fit the models, the validation set to tune hyperparameters, and the test set to assess the final performance of the models.

Non-sequential models are widely used for analyzing tabular data, where each sample is represented as a fixed-length vector and treated independently of temporal order. In our methodology, this category includes Linear Regression~\cite{linear_regression}, XGBoost~\cite{xgboost}, Random Forest~\cite{ho1995random}, and Feedforward Neural Networks (FNN)~\cite{goodfellow2016deep}, all of which are well-suited for learning from aggregated monitoring metrics and statistical features rather than raw time-series inputs. These models use a feature vector \( X = (x_1, x_2, \ldots, x_M) \), where each \( x_i \) corresponds to a feature extracted from the monitoring metrics $M$ using \texttt{tsfresh}, based on the selected observation window.

Sequential models are tailored for scenarios where the order and timing of data points carry meaningful information. This category includes Recurrent Neural Networks (RNN)~\cite{RNNs}, Long Short-Term Memory (LSTM)~\cite{hochreiter1997long}, Gated Recurrent Units (GRU)~\cite{wu2017introduction}, and Convolutional Neural Networks (CNN)~\cite{chung2014empirical}. These models are effective in handling time-series data and can learn complex temporal patterns and long-term dependencies. The input to these models is a sequence \(\textbf{X} = (X_1, X_2, \ldots, X_M)\), where each \( X_i \) represents the full time-series of a specific monitoring metric within the observation window.

The selection of candidate machine learning models can be guided by the type of correlation \( r^* \) between features of monitoring metrics~\cite{gregorutti_correlation_2017} and the target variable, as well as the size \( N \) of the dataset~\cite{10.5555/3524938.3525035}. Table~\ref{tab:combined-model-selection} presents the recommended model options for each correlation type and data size. By following these recommendations, only a selected subset of models is evaluated, which significantly reduces overall training time.

Each candidate model is evaluated based on two criteria: Root Mean Squared Error (RMSE) and inference time. Only models whose inference time remains below a defined fraction $\tau_{\text{inference}}$ of the mean RTT of the application are considered acceptable. In our work, we use $\tau_{\text{inference}} = 1\%$. Among them, the model with the lowest RMSE is selected as optimal. The optimization problem is defined as follows:

\begin{equation}
\begin{split}
    f_{\text{select}} &= \underset{ml}{\mathrm{arg\,min}} \ \text{RMSE}(ml), \\
    &\text{subject to} \quad t_{\text{inference}}(ml) \leq \mu_{\text{RTT}} \cdot \tau_{\text{inference}}
\end{split}
\label{eq:model_selection}
\end{equation}

In this formulation, \( ml \) denotes a candidate machine learning model trained on \( k^* \) input features extracted over the observation window \( w^* \). The RMSE of the model is denoted \( \text{RMSE}(ml) \), and \( t_{\text{inference}}(ml) \) is the inference time. \( \mu_{\text{RTT}} \) represents the mean RTT of tasks towards the target application.

\underline{Re-training:} The re-training strategy depends on the model type. For sequential models and feedforward neural networks, online training is used to incrementally update the model as new data become available, enabling real-time adaptation without requiring full training. For all other non-sequential models (e.g., LR, XGBoost), the model is retrained on the entire dataset while retaining the previously tuned hyperparameters.

\underline{Model Evaluation:} Each newly trained or updated model is evaluated by comparing its prediction error to that of the previous model, if one exists. Specifically, the training process calculates the relative change in RMSE between the new and the previous model as follows:

\begin{equation}
\text{RMSE}_{\text{change}} = \frac{\text{RMSE}_{\text{new}} - \text{RMSE}_{\text{prev}}}{\text{RMSE}_{\text{prev}}}
\end{equation}

If this change exceeds a predefined threshold \(\theta\), the data collection process is notified to re-evaluate the correlations between monitoring metrics and RTT. Based on the updated correlations, the input dataset is revised. Once this update is complete, the training process is triggered to train a model from scratch (full training). In our experiments, we set the threshold to \(\theta = 10\%\).

\underline{Prediction process notification:}
Once a model is created or updated, the prediction process is notified. After this notification, the training process pauses and waits until a new notification is received.

\subsection{Prediction}
The process can be configured to operate either in an on-demand mode triggered through HTTP requests or in a periodic mode where predictions are made at regular intervals. In on-demand mode, predictions are always fresh but can cause higher overhead under heavy request rates, since a prediction is computed for each request. In periodic mode, predictions are precomputed and reused, reducing overhead during high load but potentially becoming slightly outdated between updates. The prediction process consists of a sequence of steps corresponding to the components in the red panel of Figure~\ref{fig:methodology}. These steps are as follows:

\underline{Model load:}
Upon notification, the newly trained model is loaded, and it is ready for predictions. 

\underline{State retrieval:}
The process retrieves the required monitoring metrics, specifically, the top $k^*$ metrics within the optimal observation window $w^*$, determined by the data collection process, from the monitoring system. 

\underline{Feature extraction:}
Depending on the model type, it may also extract time-series features from the retrieved monitoring metrics through tsfresh.

\underline{Inference:}
The model performs an inference to estimate the RTT of the upcoming task. The predicted value is stored in a database (referred to as knowledge base) for subsequent analysis. The prediction time can be divided as follows:
\begin{equation}
    t_{\text{prediction}} = t_{\text{state}} + t_{\text{feature}} + t_{\text{inference}} 
\end{equation}

\section{Experimental Setup} \label{sec:set_up}

This section presents the applications, software frameworks, compute infrastructure, and experimental workflows deployed.

\subsection{Applications}

Our experiments employ five Single Particle Analysis (SPA) applications: \textit{Upload}, \textit{MotionCor2}, \textit{FFT Mock}, \textit{gCTF}, and \textit{ctffind4}. Each application receives a POST request with an input configuration, processes the request using a specific algorithm, and returns the result. Applications are single-threaded and handle one task at a time. Their functions are as follows:

\begin{itemize}
    \item \textbf{Upload:} Transfers an image file from the client to the target worker node.
    \item \textbf{MotionCor2:} Compensates for drift and beam-induced motion during image acquisition, improving image quality~\cite{BRILOT2012630,Zheng2017MotionCor2}. It merges multiple frames into a corrected single-frame image (MRC format~\cite{crowther_mrc_1996}). Version 1.5.0 is used.
    \item \textbf{FFT Mock:} Emulates lightweight image processing workloads by performing pixel duplication, shifting, and Fast Fourier Transform (FFT)-based operations to simulate proprietary algorithms.
    \item \textbf{gCTF:} Performs GPU-accelerated real-time Contrast Transfer Function (CTF) estimation and correction~\cite{zhang_gctf_2016}, enhancing image quality for downstream analysis. We use version 1.18.
    \item \textbf{ctffind4:} A CPU-based tool (version 4.1.14) for CTF estimation~\cite{ROHOU2015216}.
\end{itemize}

Input data is acquired from a vitrified EM grid containing a protein sample, using a Glacios electron microscope equipped with a Falcon4i detector and EPU software~\cite{deng_ispa_2021}. All applications operate on MRC files, with fraction images used by \textit{Upload} and \textit{MotionCor2}, and single integrated images by the remaining applications. Fraction files are 897MB each, and integrated files are 65MB. The dataset consists of 98 images.

\subsection{Frameworks}
We use Kubernetes~\cite{Kubernetes}, an open-source container orchestration platform, to manage distributed resources and deploy applications as pods across nodes. Each pod runs one or more containers using container runtimes like Docker~\cite{Docker} or containerd~\cite{Containerd}. The master node manages the scheduling of the pod, while the worker nodes provide the actual resources of the cluster. GPU scheduling becomes available through the installation of GPU plugins.

Monitoring is handled by Prometheus~\cite{Prometheus}, a time-series data collection system integrated with Kubernetes. The Prometheus server scrapes metrics from exporters running on nodes and containers at 200ms intervals and stores the data in a local database. Users can visualize metrics via Grafana~\cite{Grafana}, its web UI, or its HTTP API. GPU metrics are retrieved using the NVIDIA DCGM exporter~\cite{dgcm}.

Jaeger~\cite{jaegertracing} is deployed for distributed tracing, tracking end-to-end request latencies across services. It captures RTT within the cluster but excludes external delays prior to the gateway.


\subsection{Compute Infrastructure}

\begin{table}
\caption{Specifications of infrastructure components. Worker-1 is equipped with a Tesla K20c GPU, Worker-2 with a Tesla M40, and Worker-3 with an RTX 4090. All other nodes do not have a GPU. Workers 1, 2, 5, and 9–29 are equipped with HDD, while Workers 3, 4, 6, 7, and 8 use SSD.}
\begin{center}
\footnotesize
\begin{tabular}{llcc}
\toprule
\textbf{Name} & \textbf{Model/Processor} & \textbf{Cores} & \textbf{RAM} \\
\midrule
Switch & Dell PowerConnect 5448 & - & - \\
Routing & Intel Xeon X3430  & 3 & 4GB  \\
Master & Intel Xeon Silver 4109T & 8 & 32GB  \\
Monitor & Intel Xeon Silver 4109T & 8 & 78GB  \\
Worker-1 & Intel Core i7-7700 & 4 & 32GB  \\
Worker-2 & 2xIntel Xeon E5-2637 & 4 & 24GB  \\
Worker-3 & Intel Core i9-14900K & 32 & 128GB  \\
Worker-4 & Intel Core i5-9600 & 4 & 12GB  \\
Worker-5 & 2xIntel Xeon E5504 & 8 & 24GB  \\
Worker-6 & Intel Xeon Silver 4209T & 8 & 24GB  \\
Worker-7 & Intel Xeon Gold 6138T & 16 & 64GB  \\
Worker-8 & Intel Xeon Silver 4209T & 8 & 24GB  \\
Worker-9-29 & Intel Xeon X3430 & 4 & 4/8GB \\
\bottomrule
\end{tabular}
\label{tab:specifications}
\end{center}
\end{table}

\begin{figure}
\centerline{\includegraphics[width=0.5\columnwidth]{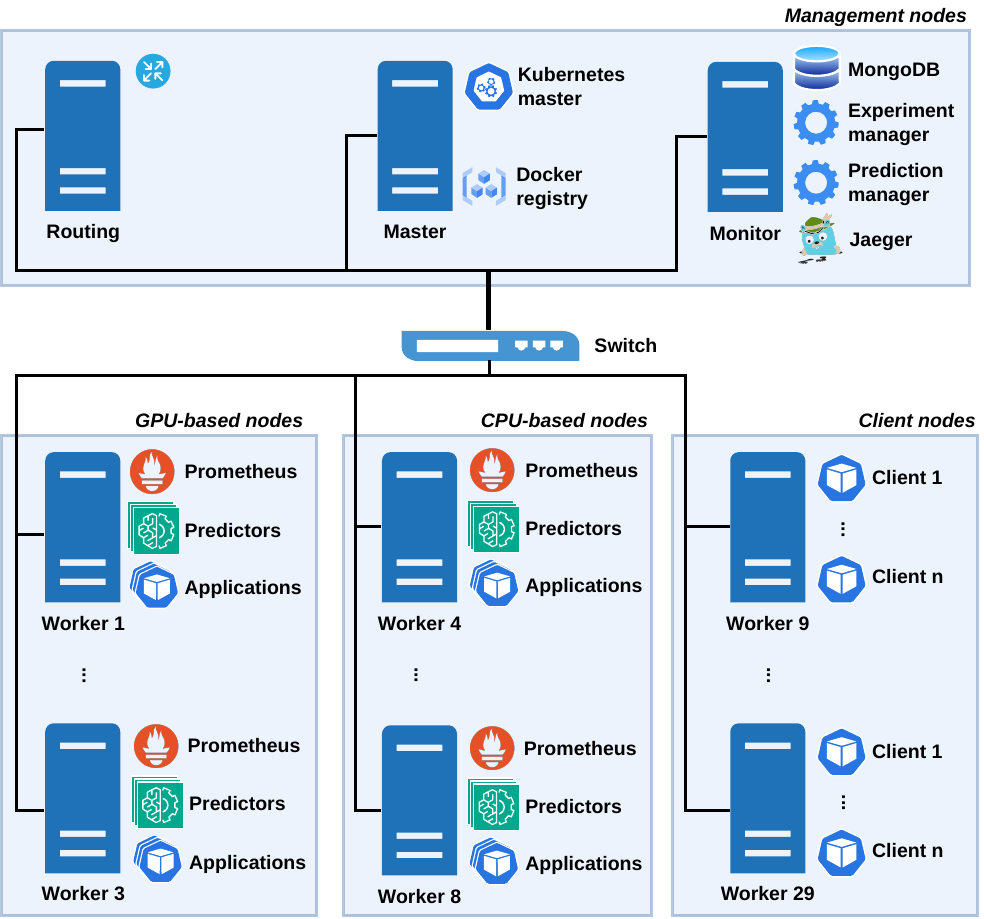}}
\caption{Cluster architecture. All nodes are interconnected via 1 Gbps Ethernet.}  
\label{fig:cluster_structure_dc}
\end{figure}

The experimental cluster (Figure~\ref{fig:cluster_structure_dc}) includes 32 interconnected servers (see Table~\ref{tab:specifications}) linked via 1 Gbps Ethernet. The Kubernetes cluster comprises one master and 30 worker nodes. As performance predictors are tailored per node, overall cluster size does not impact model accuracy. Node roles are as follows:

\begin{itemize}
    \item \textbf{Routing:} Manages internal network traffic. Not part of the Kubernetes cluster.
    \item \textbf{Master:} Hosts Kubernetes control components and the Docker image registry.
    \item \textbf{Monitor:} Runs the MongoDB, Jaeger, the Experiment Manager, and Prediction Manager.
    \item \textbf{Workers:} Workers 1-8 host applications, with Workers 1-3 specifically supporting GPU workloads. Each of these workers (1-8) also hosts its own Prometheus servers and exporters to minimize state retrieval time and eliminate network overhead during state retrieval. In addition, they host RTT predictors for the applications running on those nodes. Workers 9-29 function as clients, generating requests.
\end{itemize}

\subsection{Experimental Setup}
In our experimental setup, we deploy as many instances of each application as possible across Workers 1 to 8, until the available resources on each node are fully utilized. Resource allocation is unrestricted: applications use minimal resources while idle but can utilize as many resources as needed during active execution.  To systematically study the effects of resource contention, we gradually increase the number of co-located application instances on each node until the node becomes saturated. Once full contention is reached, we progressively reduce the load by stopping task generation for individual applications. To ensure clarity and reproducibility, the experiment is structured into distinct workload stages, each representing a specific level of resource contention. 

A \textit{workload stage} is defined as a period during which a specific subset of application instances is active. Each application has a corresponding client pod that sends a request, waits for the response, and then pauses for a random interval \( t_{\text{wait}} \sim \mathcal{U}(0, t_{\text{max}}) \) (i.e., uniformly sampled from the interval \([0, t_{\text{max}}]\)) before submitting the next request. The maximum waiting times, $t_{max}$, are set to 40, 6, 20, 10, and 10 seconds for \textit{upload}, \textit{ctffind4}, \textit{FFT Mock}, \textit{gCTF}, and \textit{MotionCor2}, respectively. These values were determined through experiments to ensure that the applications produce tasks frequently enough to induce resource contention.

The \textit{Experiment Manager} is deployed as a pod and coordinates the execution of the experiment. It uses the Kubernetes API to deploy application and client pods across all worker nodes. A separate process is initiated per node to activate the initial workload stage. In total, we run 15 workload stages, each lasting a random duration between 11 and 12 hours. Once this condition is met, the next workload stage is activated by enabling more application instances. After the final stage has completed, all application and client pods are removed.

The resources allocated to each deployed RTT predictor are limited by assigning a maximum of 10\% to the node CPU, with an upper limit of one CPU core per predictor. Restricting the CPU allocation in this way helps minimize interference with the applications running on the node. However, this limitation can also slow down the prediction processes, especially model training.

RTT predictors learn the relationship between monitoring metrics and observed RTTs. For accurate training, especially at deployment, they require diverse RTT values typically caused by resource contention. To induce this variability, the Prediction Manager temporarily adds controlled load: it deploys a \textit{noisy server} pod on the target node and a \textit{noisy client} pod on a separate worker node. The client generates light network traffic to the server, which simulates activity across key system resources: CPU, GPU, RAM, disk, and network, thus inducing controlled interference. Once all RTT predictors on the node establish the correlations between metrics and RTT, the noisy server and client are removed. RTT predictors generate predictions at regular intervals, every 5 seconds, in order to measure end-to-end prediction latency and detect any performance bottlenecks within the prediction workflow.
%
\section{Evaluation of predictors} \label{sec:predictor_results}
In this section, we present the accuracy and overhead of our RTT predictors.

\subsection{Selected correlation algorithms}
\begin{figure*}
\centerline{\includegraphics[width=1\columnwidth]{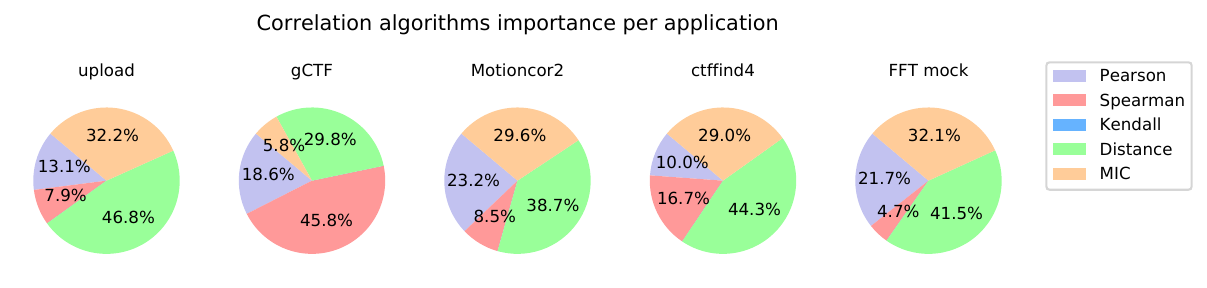}}
\caption{Proportion of monitoring metrics for each application for which each correlation method yields the highest correlation with RTT.}  
\label{fig:correlations}
\end{figure*}
Figure~\ref{fig:correlations} shows how effective the five correlation algorithms are in capturing the relationship between RTT and monitoring metrics. The algorithms compared are Pearson, Spearman, Kendall, Distance Correlation, and MIC. For each metric, we find which algorithm gives the highest correlation value. Then, for each application, we calculate the proportion of metrics for which each algorithm performs best, averaging this proportion across all nodes and instances of the application. In this way, we estimate the importance of each correlation algorithm for each application.

The results indicate that no single correlation method is optimal for all applications; instead, each application exhibits a distinct pattern reflecting its unique resource usage and runtime behavior. For example, for the \textit{upload} application, Distance Correlation (46.8\%) and MIC (32.2\%) most frequently identify the strongest relationships, indicating the presence of complex, non-linear dependencies. In contrast, for \textit{gCTF}, Spearman is most effective (45.8\%), suggesting predominantly monotonic relationships. Similarly, \textit{Motioncor2}, \textit{ctffind4}, and \textit{FFT mock} all show high proportions for Distance Correlation and MIC, further highlighting the importance of non-linear relationships in these applications. Kendall was not selected as the top method for any metric in any application instance, likely because its correlation values are similar to, but generally lower than, those of Spearman.
\subsection{Selected predictor configuration}
\begin{figure}
\centerline{\includegraphics[width=0.5\columnwidth]{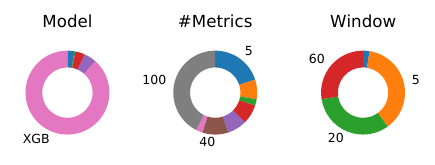}}
\caption{Proportion of selected RTT predictor configurations by model type, number of metrics, and observation window (in seconds) across all applications and nodes.}  
\label{fig:model_configuration}
\end{figure}

Figure~\ref{fig:model_configuration} shows the proportion of RTT predictor configurations that include the model type, the number of input metrics, and the observation window length across all applications and nodes. Each pie chart (from left to right) summarizes the relative frequency of selections for model type, metric count, and window size. For clarity, only options representing at least 10\% of the total selections are labeled.

The selection of the machine learning model, number of metrics, and observation window length varies across applications, highlighting that no single predictor configuration is optimal for all scenarios. Our methodology systematically explores different combinations to identify configurations that balance prediction accuracy with practical constraints, such as maintaining the prediction delay below 10\% of the RTT. This approach ensures that predictors are both accurate and efficient in real-world deployments (see Section~\ref{sec:predictor_architecture} for details on the selection process).

This process involves a trade-off: using more metrics or longer observation windows can improve prediction accuracy by providing richer system information, but also increases computational and data retrieval overhead. In contrast, reducing these parameters minimizes overhead but may affect the quality of the prediction. For instance, XGBoost is frequently selected because it offers a good balance between accuracy and speed. However, other models may be chosen if they provide higher accuracy for a specific application or better satisfy strict timing constraints.
%
\subsection{Prediction adaptation on dynamic co-location}
\begin{figure*}
\centerline{\includegraphics[width=1\columnwidth]{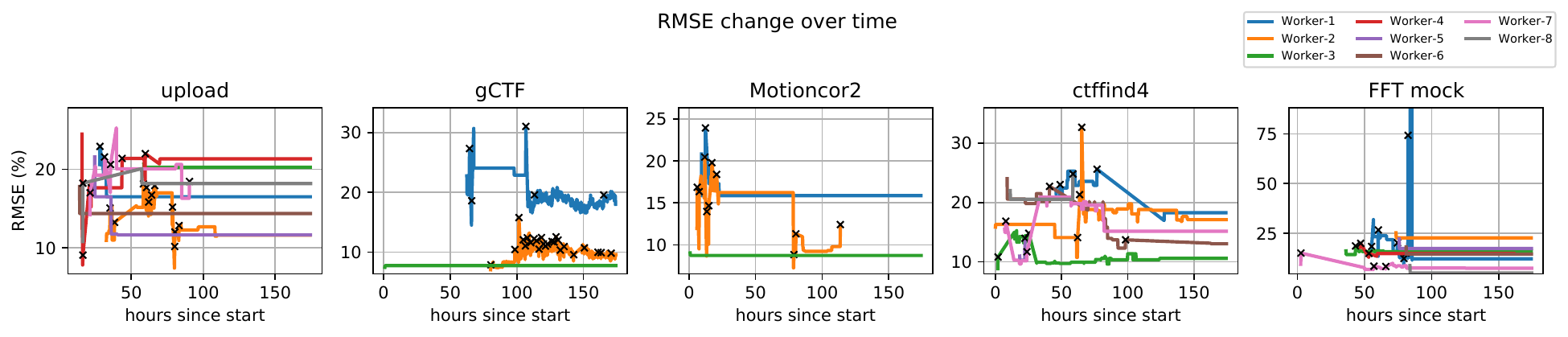}}
\caption{Evolution of RMSE over time for RTT predictors under changing co-location conditions. Each line represents a predictor deployed on a specific node. Predictors are initialized once enough training data are available, and full training events are marked along the lines.}
\label{fig:colocation_adaption}
\end{figure*}

To evaluate the accuracy and adaptability of our RTT predictors, we analyzed their RMSE over time under evolving co-location scenarios. Figure~\ref{fig:colocation_adaption} shows how the RMSE evolves for the predictors of each application as they adapt to the changing co-location patterns triggered by the arrival of new data. Each line in the plot corresponds to a predictor instance running on a specific node: there are 8 predictors for upload, ctffind4, and FFT mock, and 3 predictors for gCTF and Motioncor2. If no model satisfies the prediction delay constraint (i.e., inference time below 10\% of the RTT), the corresponding line is absent. For example, no valid predictor is deployed for ctffind4 in Worker-4. The x-axis represents the number of hours since the start of the experiment, and the y-axis shows the RMSE of each predictor. The cross markers (x) on the lines show the points where full model retraining is initiated.

We observe several clear patterns in how the RMSE changes over time. Predictors are initialized at different moments, depending on when each application starts running and when sufficient training data become available. In some cases, the RMSE increases after a training (full, online or re-training). This typically occurs because the new data introduce more complex or harder-to-predict relationships between monitoring metrics and RTT. As the system encounters more diverse co-location scenarios, the RMSE values generally stabilize, indicating that the predictor has learned the dominant interference patterns. After this point, adding more data does not significantly improve the accuracy of the model.

A sharp increase in RMSE (greater than 10\%) triggers a full training of the model from scratch. This mechanism appears in many predictors and often leads to improved accuracy. For example, in the FFT mock application running in Worker-1, the RMSE rises sharply to 550\% approximately 90 hours after the experiment begins, likely due to abrupt changes in resource usage by other applications. After full training, the new model reduces the error back to the previous levels.

However, there are exceptions. For instance, the gCTF predictor in Worker-2 initiates full training several times within a short period but sees little or no improvement in RMSE. This indicates that the lack of improvement is not due to model degradation, but rather to increased complexity and variability in the co-location scenarios, which make RTT prediction inherently more difficult. In these cases, even full training does not yield a better model because the limitation lies in the unpredictable or highly variable nature of the data itself, rather than the model. This highlights the need for a more robust strategy to decide when full training should be performed, potentially by considering the characteristics of new data instead of relying solely on error thresholds.

\begin{table}[!t]
\caption{RMSE (\%) for all app/node pairs at the final workload stage. The first column shows the worker node IDs. Missing values (“--”) indicate that no efficient predictor was found.}
\label{tab:rmse_all}
\centering
\footnotesize
\begin{tabular}{lccccc}
\toprule
        & Upload & gCTF & Motioncor2 & ctffind4 & FFT mock \\
\midrule
1 & 16.52 & 18.02 & 15.86 & 18.26 & 12.29 \\
2 & 11.64 & 9.59  & 12.41 & 17.10 & 22.74 \\
3 & 20.27 & 7.75  & 8.74  & 10.57 & 16.22 \\
4 & 21.35 & --    & --    & --    & 14.89 \\
5 & 11.66 & --    & --    & 13.97 & 17.44 \\
6 & 14.38 & --    & --    & 13.03 & 14.66 \\
7 & 18.46 & --    & --    & 15.14 & 7.54  \\
8 & 18.19 & --    & --    & 24.79 & 5.18  \\
\bottomrule
\end{tabular}
\end{table}

Table~\ref{tab:rmse_all} reports the RMSE of each RTT predictor at the end of the experiment, across all application and node pairs. The results indicate that RTT predictors generally achieve low to moderate error rates for a diverse set of applications and nodes. Applications such as \textit{upload}, \textit{gCTF}, and \textit{Motioncor2} consistently yield RMSE values below 20\%, demonstrating strong prediction accuracy. In particular, \textit{gCTF} achieves an RMSE of 7.75\% on Worker-3 and remains below 10\% on Worker-2. Similarly, \textit{Motioncor2} achieves its lowest RMSE on Worker-3 (8.74\%) and performs well on other nodes. The \textit{upload} application shows RMSE values ranging from 11.64\% to 21.35\%, indicating moderate prediction errors across all nodes.

In contrast, RMSE for \textit{ctffind4} and \textit{FFT mock} exhibit greater variability between nodes. For example, \textit{ctffind4} reaches an RMSE of 24.79\% on Worker-8, while \textit{FFT mock} peaks at 22.74\% on Worker-2. Despite these exceptions, most RMSE values remain below 20\%, confirming that the proposed predictors generalize well across different workloads and deployment environments, even in the presence of application- and node-specific differences. A more detailed analysis of how these errors impact load balancing decisions is provided in Section~\ref{sec:simulation_results}.

\subsection{Resource consumption of predictors}
\begin{figure}
\centerline{\includegraphics[width=0.4\columnwidth]{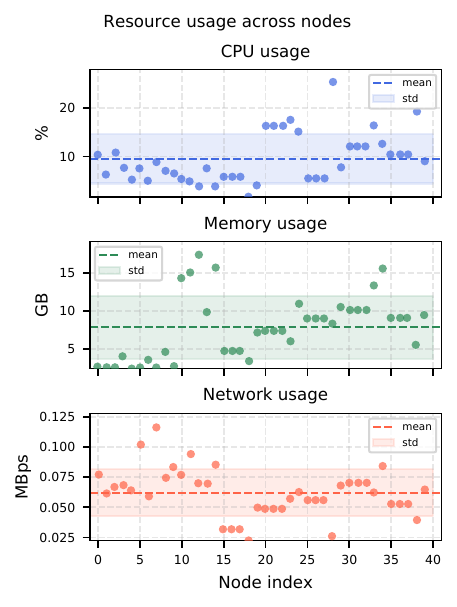}}
\caption{CPU, memory and network consumption of predictors across nodes and applications.}  
\label{fig:resource_usage}
\end{figure}

The developed prediction modules introduce resource overhead due to data collection, correlation analysis, model training, and inference. Therefore, understanding their run-time resource consumption is crucial to assess their practical feasibility. Predictors that consume excessive resources can limit the capacity available to co-located applications, reducing their suitability for deployment, even if they deliver high accuracy.

Figure~\ref{fig:resource_usage} presents the resource footprint of the predictor processes across all nodes and applications. For each predictor instance, resource usage is calculated as the average consumption measured during its operation. The figure consists of three panels, each displaying the mean and standard deviation of one resource type: CPU usage (in \%), memory usage (in GB), and network usage (in Mbps). The x-axis represents the index of each predictor instance, allowing comparison across all predictor deployments.

We observe the following:
\begin{itemize}
    \item \textbf{CPU Usage:} Predictor processes show consistently low CPU utilization, with most instances averaging below 10\% and standard deviation remaining small. The peak CPU usage does not exceed 25\%, indicating low computational burden. As described in Section~\ref{sec:set_up}, predictor processes were restricted to a single CPU core. If this restriction were relaxed, CPU usage could increase, especially for operations such as feature extraction that can benefit from parallel processing.
    
    \item \textbf{Memory Usage:} Memory consumption varies across predictors, ranging from approximately 2~GB to 18~GB. This variability is due to differences in observation window sizes, the number of stored monitoring metrics, and application RTTs. Despite these differences, all instances remain within the memory limits of modern edge nodes. The memory requirements are mainly controlled by the dataset balancing process, which limits the dataset size and helps keep memory usage low.
    
    \item \textbf{Network Usage:} Network overhead is extremely low, with mean usage consistently below 0.08~Mbps and negligible standard deviation. This efficiency is mainly achieved by deploying a local monitoring server on each node, which eliminates the need to retrieve monitoring metrics over the network. Additionally, transmitting predictions requires minimal bandwidth.
\end{itemize}

Overall, the predictors demonstrate a lightweight resource profile. The low CPU and network demands minimize interference with co-located applications, while memory usage, although variable, remains within feasible bounds. These characteristics make the proposed prediction approach well-suited for deployment in resource-constrained edge environments.

\begin{figure}
\centerline{\includegraphics[width=0.4\columnwidth]{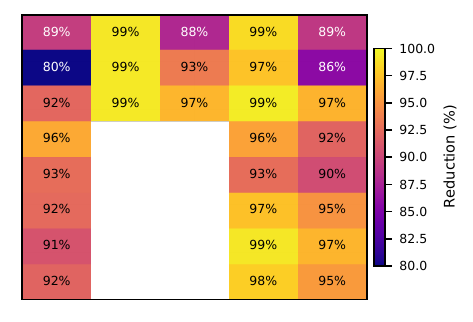}}
\caption{Dataset size reduction using dynamic bin-based sampling for each application and node pair. Each cell shows the percentage of original samples removed while maintaining RTT distribution coverage.} 
\label{fig:dataset_reduction}
\end{figure}

The storage overhead of the prediction workflow is primarily determined by the size of the collected dataset, which consists of the RTT values and the corresponding monitoring metrics from the historical tasks of each application. In our experiments, the final dataset sizes ranged from 381 to 9493 samples. To minimize storage requirements, our aim is to maintain a uniform RTT distribution in each dataset, as explained in Section~\ref{sec:predictor_architecture}.

Figure~\ref{fig:dataset_reduction} illustrates the effectiveness of our dataset reduction method. The plot presents the percentage of samples removed across various application and node combinations, with reduction rates typically between 85\% and 99\%. Although predictors were not trained on the full original datasets, due to their dataset size and memory constraints on small nodes, models built from the reduced datasets offer a favorable balance between prediction accuracy and resource overhead. Using the complete datasets would significantly increase both memory usage and training time, making them impractical for our environment. The dynamic sampling method compacts the dataset by eliminating redundant RTT values while preserving less common ones, thus maintaining RTT diversity essential for effective predictor training without unnecessary dataset growth.

\subsection{Prediction time breakdown}
\begin{figure}
\centerline{\includegraphics[width=0.4\columnwidth]{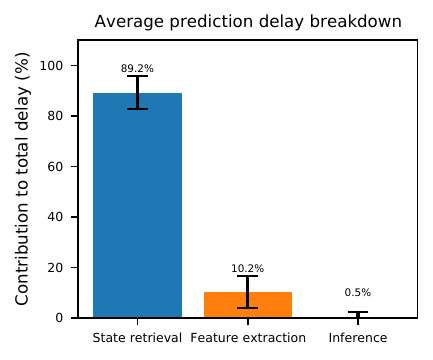}}
\caption{Contribution of state retrieval, feature extraction, and inference to the total prediction time.}  
\label{fig:prediction_delay_breakdown}
\end{figure}

The total prediction time is critical for maintaining low RTT and consistent performance in time-sensitive applications. When a prediction is triggered, the overall delay consists of three main stages: \textit{state retrieval}, \textit{feature extraction}, and \textit{inference}. Figure~\ref{fig:prediction_delay_breakdown} shows the average percentage contribution of each stage to the total prediction time across all application-node pairs, with values normalized to the RTT for direct comparison.

The results indicate that state retrieval is the dominant contributor, accounting for 89.2\% of the total delay. Feature extraction adds 10.2\%, while inference time is minimal at 0.5\%. This reveals that the primary bottleneck lies in monitoring data retrieval (e.g., via Prometheus). Achieving millisecond-scale RTT predictions will therefore require a much faster monitoring system.

Although Prometheus is deployed locally on each node, retrieval delays remain high. Further reductions may be achieved by using alternative monitoring systems or by accessing metrics directly from exporters. Alternative systems for faster metric access include VictoriaMetrics~\cite{victoriametrics} and InfluxDB~\cite{influxdb} (with Telegraf as an external collector). This limitation stems from the design of Prometheus, which is optimized for system-wide visibility rather than real-time, sub-second data retrieval. However, addressing this bottleneck falls within the domain of monitoring system optimization rather than the design of the RTT predictor itself.
\subsection{State retrieval and feature extraction overhead}
\begin{figure}
\centerline{\includegraphics[width=0.4\columnwidth]{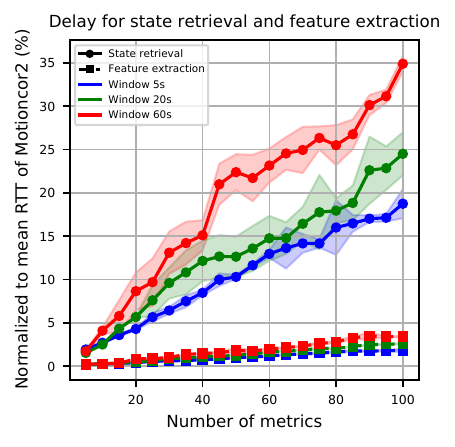}}
\caption{State retrieval and feature extraction overhead for different observation windows and number of metrics. The delays are normalized to the mean RTT of Motioncor2 in Worker-3.}  
\label{fig:delay_different_windows}
\end{figure}

Since the total prediction time is mainly dominated by the \textit{state retrieval} and \textit{feature extraction} stages, we further analyze how these delays vary with different state sizes. The state size is defined by the observation window length and the number of monitoring metrics. Figure~\ref{fig:delay_different_windows} illustrates the delay introduced by \textit{state retrieval} (from Prometheus) and \textit{feature extraction} (using \texttt{tsfresh}). These delays are presented for the case of \textit{Motioncor2} running on Worker-3 to make it easier to interpret the results and compare different combinations of state and observation window.

We evaluate three observation window sizes—5, 20, and 60 seconds—and different numbers of input metrics, ranging from 5 to 100. Each setup is executed five times to capture variability in execution time while avoiding excessive load on the monitoring system. The x-axis represents the number of metrics, and the y-axis shows the delay as a percentage of the mean RTT of Motioncor2 on Worker-3. Solid lines indicate the average delay, and shaded regions represent the standard deviation.

The results reveal distinct trends between the two stages. Feature extraction consistently incurs a low delay in all combinations of state and observation window, gradually increasing with the number of metrics but remaining below 5\% of the RTT even in the worst case. In contrast, the state retrieval delay increases significantly with both the number of metrics and the observation window length. For a 5-second window, the state retrieval delay remains below 20\% of the RTT, even when using 100 metrics. However, for a 60-second window, the delay increases sharply, reaching 35\% of the RTT at 100 metrics. These results show that feature extraction is efficient and continues to perform well even as the number of metrics and length of observation windows increases. In contrast, state retrieval becomes the main source of delay when using a larger number of metrics or longer observation windows.


This analysis demonstrates a clear trade-off between observation window size, number of metrics, and total prediction delay. Shorter windows and fewer metrics reduce overhead, making them preferable for real-time applications. In contrast, longer windows and larger metric sets provide more context but introduce significant delays, mainly due to the time required to query Prometheus. These findings further confirm that Prometheus is the dominant bottleneck in the prediction pipeline, especially for configurations that involve high-volume metric collection over extended periods. Consequently, some setups become impractical for real-time use due to high retrieval delays. For example, no machine learning model was trained for \textit{ctffind4} on \textit{Worker-4}, since the prediction delay exceeded the acceptable threshold for that application.
\subsection{Predictor deployment and RTT variability}
\begin{table*}[]
\caption{Coefficient of Variation (CoV) of application RTT for all application–node pairs, comparing scenarios with and without predictor deployment.}
\begin{center}
\scriptsize
\begin{tabular}{cc c c c c c c c c c }
\toprule
\multirow{2}{*}{\textbf{Node}}  & \multicolumn{2}{c}{\textbf{upload}} & \multicolumn{2}{c}{\textbf{gCTF}} & \multicolumn{2}{c}{\textbf{Motioncor2}} & \multicolumn{2}{c}{\textbf{ctffind4}} & \multicolumn{2}{c}{\textbf{FFT mock}} \\
\cline{2-11}
& \textbf{With} & \textbf{Without} & \textbf{With} & \textbf{Without} & \textbf{With} & \textbf{Without} & \textbf{With} & \textbf{Without} & \textbf{With} & \textbf{Without} \\
\midrule
\textbf{Worker-1} & 37.92\% & 36.43\% & 48.64\% & 50.81\% & 43.61\% & 41.70\% & 70.23\% & 17.04\% & 3.78\% & 3.82\% \\ 
\textbf{Worker-2} & 49.51\% & 52.04\% & 45.11\% & 36.62\% & 48.15\% & 49.52\% & 27.18\% & 18.53\% & 8.84\% & 9.12\% \\ 
\textbf{Worker-3} & 21.31\% & 21.46\% & 20.33\% & 9.86\% & 32.39\% & 14.36\% & 9.37\% & 9.47\% & 8.47\% & 8.38\% \\ 
\textbf{Worker-4} & 21.74\% & 17.88\% & - & - & - & - & 23.41\% & 15.54\% & 5.56\% & 3.48\% \\ 
\textbf{Worker-5} & 21.62\% & 22.94\% & - & - & - & - & 28.61\% & 8.97\% & 5.40\% & 4.40\% \\ 
\textbf{Worker-6} & 17.27\% & 16.93\% & - & - & - & - & 9.63\% & 8.47\% & 7.27\% & 6.79\% \\ 
\textbf{Worker-7} & 19.85\% & 21.20\% & - & - & - & - & 8.66\% & 8.35\% & 9.53\% & 11.32\% \\ 
\textbf{Worker-8} & 20.82\% & 20.58\% & - & - & - & - & 8.14\% & 8.30\% & 9.36\% & 7.93\% \\ 
\bottomrule
\end{tabular}
\label{tab:cov_all}
\end{center}
\end{table*}

We evaluate the impact of running RTT predictors alongside target applications on the same worker nodes, where both share CPU and memory resources. To quantify this effect, we repeat each experiment twice: once with predictors active and once without, using identical co-location scenarios and durations. We then compare the RTT of application tasks between the two scenarios. Table~\ref{tab:cov_all} reports the Coefficient of Variation (CoV) for each application/node pair, which quantifies RTT variability relative to its mean and is defined as:
\begin{equation}
CoV_{\mathrm{RTT}} = \frac{\sigma_{\mathrm{RTT}}}{\mu_{\mathrm{RTT}}}
\label{cov}
\end{equation}
where $\sigma_{\mathrm{RTT}}$ and $\mu_{\mathrm{RTT}}$ denote the standard deviation and mean RTT, respectively. A higher CoV indicates greater performance variability.

The results reveal mixed effects on application performance. In some cases, deploying predictors increases RTT variability, while in others the CoV remains stable or decreases slightly. For example, on Worker-1, the CoV for \textit{ctffind4} increases sharply from 17.04\% to 70.23\%, indicating significant resource contention introduced by the predictor. Conversely, \textit{FFT mock} on the same node shows negligible change (3.78\% vs.\ 3.82\%). On Worker-2, \textit{ctffind4} experiences higher variability (27.18\% vs.\ 18.53\%), while \textit{FFT mock} sees a slight reduction (8.84\% vs.\ 9.12\%). Worker-3 exhibits notable increases in CoV for applications \textit{Motioncor2} (32.39\% vs.\ 14.36\%) and \textit{gCTF} (20.33\% vs.\ 9.86\%). In contrast, nodes like Worker-7 and Worker-8 display minimal changes.

These differences are primarily attributed to the additional resource usage by predictors. Although operations like feature extraction and model training are usually lightweight and do not run often, they can occasionally cause short bursts of high CPU and memory usage. During these bursts, especially when the node is already running multiple applications, less CPU and memory are available for the main application, which can temporarily affect its performance. Even minor increases in resource usage can affect sensitive applications. Occasional decreases in CoV when predictors are present are likely due to random fluctuations or measurement noise, rather than a true performance improvement.

Overall, running predictors on the same node as applications can introduce additional runtime variability, with the extent depending on the specific node and application sensitivity to resource contention. The predictors are designed to capture their own resource usage through monitoring metrics, allowing them to maintain valid RTT predictions even during concurrent operation. However, for production environments demanding minimal performance variability, alternative deployment strategies may be needed. These strategies include offloading predictors to dedicated nodes, scheduling intensive predictor tasks during low-load periods, or reserving extra CPU and memory on shared nodes. Reserving resources guarantees that the application has exclusive access to a portion of the system capacity, thereby minimizing performance drops caused by competition with predictor processes.
\section{Impact on load-balancing} \label{sec:simulation_results}
To analyze the impact of performance\allowbreak -aware load balancing compared to other strategies in edge computing environments, we developed a simulation framework that models a cluster of heterogeneous nodes running multiple application replicas. The simulation evaluates how placement decisions affect RTT and resource utilization under varying hardware configurations and co-location scenarios. All reported results are averaged over 200 independent trials to ensure statistical reliability.

\subsection{Simulation Workflow}
Each simulation trial begins by initializing a set of compute nodes (for example, 10 nodes), where each node is assigned a number of CPU cores and a memory capacity randomly selected from predefined ranges to reflect hardware heterogeneity. To capture additional variation in hardware performance, each node is also assigned an acceleration factor. Applications are characterized by predefined mean RTT, resource requirements, and sensitivity to performance interference. All simulation parameters are derived from measurements collected on a real-world Kubernetes cluster running SPA workloads. Due to space constraints, these parameters are not shown here, but are available in the public repository (see the corresponding section). The goal of the simulation is to emulate realistic system dynamics based on these measurement-driven parameters, rather than to predict absolute performance values. 

Interference effects are modeled using an empirically derived interference matrix, which quantifies the increase in RTT when applications are co-located on the same node. The placement of replicas across nodes is randomized in each trial to isolate the impact of the load-balancing strategy.

For each request, the RTT is computed as:
\begin{equation}
    \text{RTT}_{\text{actual}} = x \cdot (1 + \alpha), \quad \text{where } x \sim \text{LogNormal}(\mu, \sigma),
\end{equation}
where $\alpha$ is the acceleration factor. The log-normal parameters $\mu$ and $\sigma$ are derived using the application mean RTT $\bar{r}$ (fixed for each application) and the standard deviation $s$, which is determined for each co-location scenario using an empirically constructed interference matrix. This matrix quantifies the increase in RTT variability due to resource contention between different applications running on the same node. The log-normal distribution was selected based on RTT measurements from our experiments. The values of $\mu$ and $\sigma$ are given by:
\begin{equation}
    \mu = \log\left(\frac{\bar{r}^2}{\sqrt{s^2 + \bar{r}^2}}\right), \qquad
    \sigma = \sqrt{\log\left(1 + \frac{s^2}{\bar{r}^2}\right)}
    \label{eq:lognorm_params}
\end{equation}

Three load balancing strategies are considered:
\begin{itemize}
    \item \textbf{Round Robin:} Replicas are selected in sequential order if idle.
    \item \textbf{Random Selection:} Replicas are chosen randomly from available idle replicas.
    \item \textbf{Performance\allowbreak -Aware Load Balancing:} Idle replicas are evaluated based on predicted RTTs. Each predicted RTT is computed as a noisy estimate of the true RTT:
    \begin{equation}
        \text{RTT}_{\text{predicted}} = \text{RTT}_{\text{actual}} + \mathcal{N}(0, \epsilon),
    \end{equation}
    where \( \epsilon = (1 - p) \cdot \text{RTT}_{\text{actual}} \) and \( p \in [0, 1] \) is the prediction accuracy. The replica with the lowest predicted RTT is selected.
\end{itemize}
Requests are scheduled according to the chosen strategy, and each replica maintains a local busy-until timestamp to enforce concurrency limits. For each request, the total CPU and memory used by a replica is calculated by multiplying its per-request resource requirement by the time the request takes to finish.

\subsection{Expected gain in load-balancing}
\begin{figure*}
\centerline{\includegraphics[width=1\columnwidth]{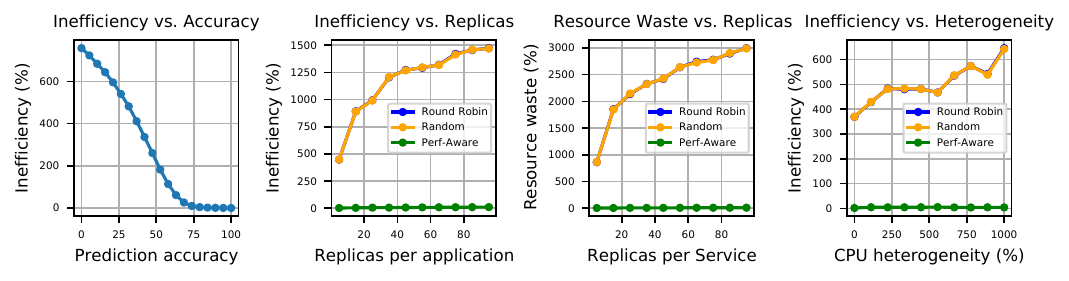}}
\caption{Simulation results of RTT predictor integration into load balancing.}  
\label{fig:simulation}
\end{figure*}

We compare the performance\allowbreak -aware load balancing strategy against traditional methods such as round-robin and random assignment. The evaluation focuses on four key factors: prediction accuracy, number of replicas, resource waste, and system heterogeneity. To quantify the effectiveness of the load balancer, we use the metric \textit{scheduling inefficiency}, defined as the performance loss (in terms of RTT or resource usage) relative to an ideal load balancer with perfect knowledge of system performance.

The effectiveness of the performance\allowbreak -aware load balancer depends on the accuracy of its performance predictions. The first subplot of Figure~\ref{fig:simulation} illustrates how scheduling inefficiency changes with prediction accuracy. As prediction accuracy increases, scheduling inefficiency decreases, dropping to nearly zero when accuracy reaches 80\%. This shows that a near-perfect prediction is not required for efficient load balancing. Beyond this threshold, further increases in accuracy do not lead to further gains, as the load balancer is already able to consistently select faster application instances and avoid slower ones. In contrast, lower prediction accuracy leads to more suboptimal placement decisions and higher inefficiency. The accuracy threshold for optimal performance may vary depending on cluster characteristics, such as hardware heterogeneity. For all subsequent simulation results, we assume that the predictors achieve 80\% accuracy.

The second subplot of Figure~\ref{fig:simulation} illustrates how scheduling inefficiency changes with the number of replicas per application for different load balancing strategies. As the number of replicas increases, the performance\allowbreak -aware load balancer maintains low inefficiency, while round-robin and random strategies exhibit increasing inefficiency. This shows that performance\allowbreak -aware load balancing scales effectively by leveraging RTT predictions for task placement. In contrast, the baseline strategies ignore application interference, resulting in more suboptimal placements as the number of replicas, and possible placement options, grows.

In resource-constrained environments, minimizing both RTT and overall resource consumption is essential. The decisions made by the load balancer directly affect how efficiently cluster resources are used. The third subplot of Figure~\ref{fig:simulation} shows resource waste as the number of replicas increases for each strategy, where resource waste refers to the extra resources consumed compared to an ideal load balancer with perfect knowledge. For both round-robin and random strategies, resource waste increases as more replicas are added, since these approaches do not account for current system state or performance and often lead to inefficient placements. In contrast, the performance\allowbreak -aware load balancer keeps resource waste significantly lower and scales more efficiently by using RTT predictions to select suitable replicas, distribute workload evenly, and avoid unnecessary contention. This results in better resource utilization, as tasks are completed faster and resources are released more quickly, preventing both over-provisioning and underutilization. The simulation also accounts for any additional consumption from the RTT predictor itself.

Resource clusters typically include nodes with varying hardware and software configurations, leading to different RTTs for the same task across nodes. Hardware heterogeneity thus plays a key role in load-balancing effectiveness. The fourth subplot of Figure~\ref{fig:simulation} shows that scheduling inefficiency rises as CPU heterogeneity increases. Traditional strategies like round-robin and random assignment are less effective in this context, as they overlook node performance differences. In contrast, the performance\allowbreak -aware load balancer adapts to resource variability, maintaining low inefficiency even as heterogeneity grows. By leveraging RTT predictors, it makes informed placement decisions that account for hardware differences, resulting in more efficient and predictable task execution.

Overall, the simulation results demonstrate that performance\allowbreak -based load balancing, guided by RTT prediction, offers substantial gains in efficiency, scalability, and adaptability, especially in heterogeneous and dynamic environments. These advantages make it an attractive solution for deployment in modern cloud and edge-based systems.

\section{Lessons Learned} \label{sec:lessons}
This section highlights the key insights gained from developing and evaluating our performance prediction approach, focusing on practical considerations and trade-offs relevant to edge computing environments:
\begin{enumerate}
    \item \textbf{Diverse correlation methods:} Both linear and non-linear relationships exist between RTT and monitoring metrics. Applying a range of correlation techniques is essential to accurately identify the most relevant metrics and ensure robust predictor design.

    \item \textbf{Configuration space exploration:} No single configuration (machine learning model, metric number, observation window), performs best in all cases. Each application-node pair requires its own set of parameters to effectively balance prediction accuracy and overhead.



    \item \textbf{Adaptation and retraining limitations:} Predictors typically adapt to changing co-location patterns, with retraining helping RMSE stabilize as more data are collected. However, in highly complex scenarios, full training may not improve accuracy, since errors can result from more challenging relationships between system state and RTT, rather than from suboptimal predictor configurations.

    \item \textbf{Predictor resource usage:} The predictors impose minimal overhead, characterized by low CPU and network usage, moderate memory requirements, and reduced storage needs as a result of the proposed data balancing strategy. These properties make the approach suitable for real-time deployment in resource-constrained environments.

    \item \textbf{Prediction time bottleneck:} RTT prediction delay is dominated by monitoring data retrieval, not inference. Faster monitoring systems are needed for real-time operation.

    \item \textbf{Impact of state size:} Using larger observation windows or more metrics increases the amount of data the monitoring system must provide, which can slow down prediction. To keep predictions fast, it is essential to carefully select the window size and limit the number of metrics.

    \item \textbf{Predictor effect on variability:} Running predictors alongside applications, even with limited resources, can increase performance variability. This impact depends on the specific application and node. Interference can be minimized by using dedicated nodes for predictors or scheduling its resource-intensive operations (i.e. training) during low system load.

    \item \textbf{Prediction accuracy threshold:} Once prediction accuracy is sufficiently high, further improvements have little effect on load balancing efficiency

    \item \textbf{Awareness of interference and heterogeneity:} Performance\allowbreak -aware load balancing maintains efficiency (i.e. low RTT) as cluster complexity increases by accounting for application interference and node heterogeneity, while simple strategies become less effective.

    \item \textbf{Reducing resource waste:} Using RTT predictions allows performance\allowbreak -aware load balancing to minimize resource waste, supporting efficient scaling with more replicas.
\end{enumerate}

\section{Conclusions and future work} \label{sec:conclusions}

We propose Morpheus, a practical framework for building RTT predictors using the historical state of monitoring data. Instead of arbitrarily selecting input features, our approach leverages the previously developed perfCorrelate framework to identify the most relevant monitoring metrics. The resulting predictors run in parallel with applications, providing fast predictions with minimal resource overhead while maintaining high accuracy under diverse runtime conditions, including resource contention and hardware heterogeneity. This is achieved by carefully balancing the construction of the dataset, the length of the observation window, the selection of features and the complexity of the model, all guided by the correlation between the monitoring metrics and the RTT.

To evaluate the impact of prediction-aware load balancing, we simulate a realistic edge computing environment where RTT predictors are integrated into the request routing process. Our results show that performance\allowbreak -aware load balancing driven by accurate RTT predictions significantly improves application performance, reduces resource waste, and enhances robustness in heterogeneous systems. Additionally, we outline key lessons learned throughout the development and evaluation process to guide future improvements.

In future work, we plan to integrate RTT predictors into a production-grade load balancer to assess their impact on real-world cluster. We will also refine the predictors to reduce delay without losing accuracy and benchmark them against state-of-the-art prediction and load balancing methods.

\section*{Acknowledgments}
This publication is part of the project ADAPTOR: Autonomous Distribution Architecture on Progressing Topologies and Optimization of Resources (with project number 18651 of the research Open Technology Programme which is (partly) financed by the Dutch Research Council (NWO). The image data used for the experiments was acquired at Thermo Fisher Scientific (Eindhoven, The Netherlands). The computational experiments, including execution of the described algorithms, were executed at the Eindhoven University of Technology (Eindhoven, The Netherlands).

\bibliographystyle{unsrt}  
\bibliography{ref}

\end{document}